%% file: proc_slacpub.tex
\documentclass[11pt]{article}
\usepackage{graphicx}
\usepackage{epsfig}

\newcommand{\BABARPubYear}    {00}

\newcommand{\BABARProcNumber} {17}
\newcommand{\SLACPubNumber} {8679}
\newcommand{\LANLNumber} {0000}

\input pubboard/babarsym
\long\def\inst#1{\par\nobreak\kern 4pt\nobreak
    {\it #1}\par\vskip 10pt plus 3pt minus 3pt}

\begin{document}
{\pagestyle{empty}

\begin{flushright}
SLAC-PUB-\SLACPubNumber \\
\babar-PROC-\BABARPubYear/\BABARProcNumber \\
hep-ex/\LANLNumber \\
August, 2000 \\
\end{flushright}

\par\vskip 4cm

\begin{center}
\Large \bf 
 Measurements of Inclusive and Exclusive \B\ Decays to Charmonium with \babar
\end{center}
\bigskip

\begin{center}
\large 
Gerhard Raven\\
University of California at San Diego\\
La Jolla, CA92093, USA\\
(for the \lbabar\ Collaboration)
\end{center}
\bigskip \bigskip

\begin{center}
\large \bf Abstract
\end{center}
   Using 8.5M \BB\ events recorded by the \babar\ detector 
   the yield of inclusive \jpsi, the branching ratios of \psitwos\ and
   \chic\ are presented. Combining the charmonium state with either 
   a \Kpm,\KS, \Kstarpm\ or \Kstarz\ exclusive B decays are reconstructed
   and their branching ratios determined. Using the fully reconstructed
   decays, both the \Bz\ and \Bpm\ masses and their difference is measured.
   Finally the contributions of \CP\ even and odd amplitudes in the decay
   \B\to\jpsi\Kstar are determined from an angular analysis.

\vfill
\begin{center}
Contributed to the Proceedings of the 30$^{th}$ International 
Conference on High Energy Physics, \\
7/27/2000---8/2/2000, Osaka, Japan
\end{center}

\vspace{1.0cm}
\begin{center}
{\em Stanford Linear Accelerator Center, Stanford University, 
Stanford, CA 94309} \\ \vspace{0.1cm}\hrule\vspace{0.1cm}
Work supported in part by Department of Energy contract DE-AC03-76SF00515.
\end{center}

\setlength\columnsep{0.20truein}
\twocolumn
\def\sloppy{\tolerance=100000\hfuzz=\maxdimen\vfuzz=\maxdimen}
\sloppy
\vbadness=12000
\hbadness=12000
\flushbottom
\def\figurebox#1#2#3{%
  	\def\arg{#3}%
  	\ifx\arg\empty
  	{\hfill\vbox{\hsize#2\hrule\hbox to #2{\vrule\hfill\vbox to #1{\hsize#2\vfill}\vrule}\hrule}\hfill}%
  	\else
   	{\hfill\epsfbox{#3}\hfill}%
  	\fi}

\section{Introduction}

Observation of charmonium mesons in \B\ decays is a crucial component of the
measurement of time-dependent \CP-violating asymmetries\cite{conf0001}. 

%
For the analyses described here, a sample of 7.7 \invfb\ collected at the \FourS\
resonance 
with an additional 1.2 \invfb\ collected below the \BB\ threshold 
are used.
The number of \BB\ events is determined by counting the number of hadronic events 
selected both on and off resonance. The continuum contribution to the on-resonance 
sample is estimated by rescaling the number of off-resonance hadronic events by the 
ratio of the number of observed \mumu\ events in the two samples. This procedure 
yields a total of $8.46 \pm 0.14 \cdot 10^6$ selected \BB\ events.

\section{Inclusive decays of \B\ to Charmonium}

Events containing a \jpsi\ are selected by requiring two identified leptons of 
opposite charge.
Electrons are selected by requiring the observed energy in the calorimeter
to match the measured momentum, the shape of the
calorimeter cluster and the observed ionization in the tracking detectors.
Muons are identified by requiring a minimum ionizing signal in the calorimeter
and by their penetration into and observed cluster shape in the instrumented flux
return.
The number of \jpsi\ events is determined by fitting the invariant mass distribution
to a pdf obtained from a simulation which includes both final state radiation and
bremsstrahlung. The fits yield $4920\pm100\pm180$ \jpsi\to\epem\ and $5490\pm90\pm90$
\jpsi\to\mumu\ signal events.

Events containing \psitwos\ decays are reconstructed in both the leptonic decays
of the \psitwos\ and its decays to \jpsi\pip\pim. In case of the former, the number
of signal events is extracted in similar fashion to the \jpsi; for the latter, a fit
to the mass difference between the \psitwos\ and the \jpsi\ candidates is
performed. We find $131 \pm 29 \pm 2$ decays to \epem,
$125 \pm 19$ to \mumu,
$126\pm44$ to \jpsi(\mumu)\pip\pim\ and
 $162\pm23$ to \jpsi(\epem)\pip\pim.

The \chic{1} and \chic{2} are reconstructed by combining a \jpsi\ candidate with a photon. The
signal yield is determined by fitting the mass difference between the \chic{ }
and \jpsi\ candidates.  We fit simultaneously for a \chic{1}\ and possible \chic{2}\ 
component. The shape of the signal is taken from the simulation, and the mass 
difference between the \chic{1} and \chic{2} is fixed to the PDG value\cite{PDG}.
We find $129\pm26\pm13$ \chic{1} and $3\pm21$ \chic{2} candidates in which \jpsi\to\epem\ and
$204\pm47\pm12$ \chic{1} and $47\pm21$ \chic{2} candidates in which \jpsi\to\mumu.

The 
branching ratios of \B\to\psitwos\Xpart\ and \B\to\chic{1}\Xpart\ are determined\cite{conf0004}
by measuring their rates relative to the measured \jpsi\ yield,
and a limit is set on the decay to \chic{2}\Xpart. The results are summarized in 
Table~\ref{tab:incbr}.
\begin{table}
\begin{center}
\caption{Measured Inclusive Branching Ratios}\label{tab:incbr}
\begin{tabular}{|ll|}
\hline
Mode &  Br ($\times 10^{-2})$ \\
\hline
\psitwos&  $0.25\pm0.02\pm0.02$\\
\chic{1}&  $0.39\pm0.04\pm0.04$\\
\chic{2}&  $<0.24$ ($90$\%CL)\\
\hline
\end{tabular}
\end{center}
\end{table}

\section{Exclusive decays of \B\ to Charmonium}

As the exclusive decays in general have very little background,
lepton identification is required for only one of the two \jpsi\
decay products.
After including  photons compatible with bremsstrahlung from one
of the leptons, the charmonium states are selected in a window
around their expected mass\cite{PDG}, and the observed momenta
are refined by a kinematic fit constraining the charmonium masses.

The charmonium candidates are then combined with either a \Kp,
a \KS\ (either \pip\pim\ or \piz\piz), \Kstarz\ (either \Kp\pim\ or
\KS\piz) or \Kstarp\ (either \KS\pip\ or \Kp\piz) to form a 
\B\ candidate.
The two most significant observables used to identify the
signal are $\Delta E$, the difference in energy between the reconstructed
\B\ decay and $\sqrt{s}/2$, and the energy-substituted \B\ mass,
$\mes = \sqrt{\left(\sqrt{s}/2 \right)^2 - {P_B^*}^2}$ where $P_B^*$ is 
the center of mass momentum of the \B\ candidate.
An example of these distributions is shown in figure~\ref{fig:mes}.
In the case of multiple candidates per event, only the candidate with the 
smallest $|\Delta E|$ is selected. 
\begin{figure*}
\epsfig{file=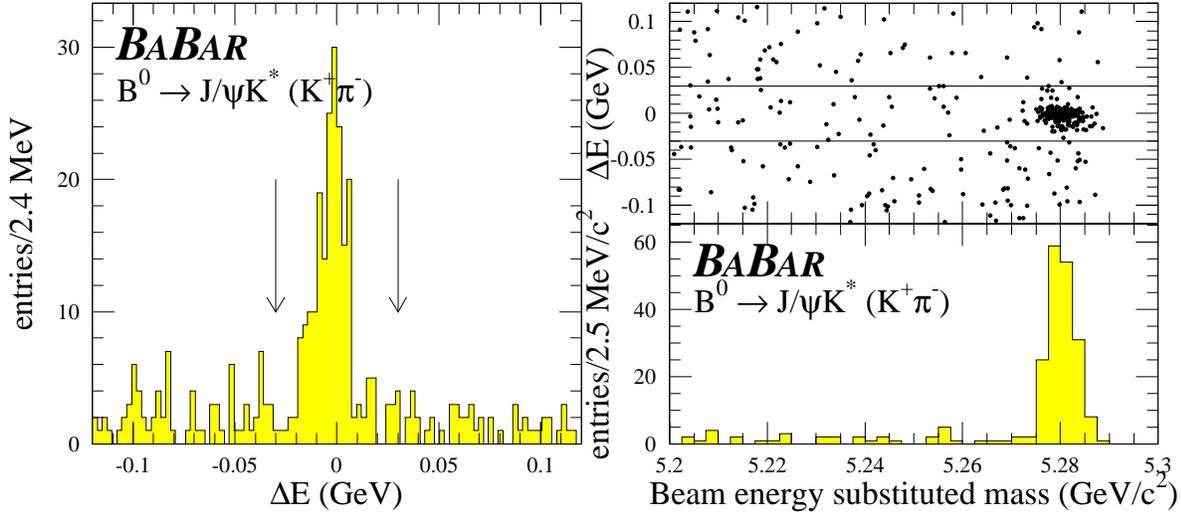,width=0.95\linewidth}
\caption{Example of the $\Delta E$ (left) and \mes\ (right) distributions
for the decay \B\to\jpsi\Kstar(\Kp\pim).}
\label{fig:mes}
\end{figure*}

The signal yields are determined 
by fitting the \mes\ distribution with the sum of a Gaussian and an 
ARGUS function\cite{Argus}; for the \Kstar\ modes a likelihood fit is
performed to all modes simultaneously, taking into account the cross-feed
between the decays.

Systematic uncertainties considered include
     the number of produced \B\ events (3.6\%),
     the signal fit (0.9--8.6\%),
     uncertainties on the measured tracking (2.5\% per track),
     neutral (0.6--11\%) and particle ID (2.5--8.8\%) efficiencies,
     the tracking resolution (0.6--2.6\%),
     the branching ratios of secondary decays (2.2--13.1\%) and
     MC statistics (0.5--5.8\%).
The observed yields and branching ratios\cite{conf0005} are summarized in Table~\ref{tab:br}.
\begin{table}
\begin{center}
\caption{Measured Exclusive Branching Ratios}\label{tab:br}
\begin{tabular}{|lll|}
\hline
Mode & Yield & Br ($\times 10^{-4})$\\
\hline
\Bpm\to\jpsi\Kstarpm\hspace*{-4mm}&$126\pm12$&$13.2\pm1.4\pm2.1$\\
\Bz\to\jpsi\Kstarz\hspace*{-4mm}&$188\pm14$&$13.8\pm1.1\pm1.8$\\
\Bpm\to\jpsi\Kpm\hspace*{-4mm}& $445\pm21$&$11.2\pm0.5\pm1.1$\\
\Bz\to\jpsi\Kz\hspace*{-4mm}&&\\
  $\;\;\;\;$\KS\to\pipi\hspace*{-4mm}&$\;\,93\pm10$ &$10.2\pm1.1\pm1.3 $\\
  $\;\;\;\;$\KS\to\piz\piz\hspace*{-4mm}&$\;\,14\pm4$ &$\;\,7.5\pm2.0\pm1.2 $\\
\Bpm\to\psitwos\Kpm\hspace*{-4mm}& $\;\,73\pm8$  &$\;\,6.3\pm0.7\pm1.2 $\\
\Bz\to\psitwos\Kz\hspace*{-4mm}&&\\
  $\;\;\;\;$\KS\to\pipi\hspace*{-4mm}& $\;\,23\pm5$ &$\;\,8.8\pm1.9\pm1.8 $\\
\Bpm\to\chic{1}\Kpm\hspace*{-4mm}& $\;\,44\pm9$  &$\;\,7.7\pm1.6\pm0.9 $\\
\hline
\end{tabular}
\end{center}
\end{table}

\section{Measurement of \B\ meson masses}

The masses of the \B\ mesons are measured
using fully reconstructed decays of \Bz\to\jpsi\KS(\pip\pim), \Bz\to\jpsi\Kstarz(\Kp\pim) 
and \Bpm\to\jpsi\Kpm.
These modes are chosen
for their small backgrounds and good knowledge of the masses of their 
decays products. 

The invariant mass of the \B\ candidates is derived by fitting the decay products
to a common vertex, constraining the \jpsi\ and \KS\ masses to their nominal values.
Uncertainties in the magnetic field and the alignment of the tracking detectors could 
introduce a bias in the momentum
measurement. Their effect is quantified by comparing
the reconstructed \jpsi\ and \KS\ masses with the PDG values\cite{PDG}. 
The effect of background on the measurement has been estimated by removing
separately the $N$ events with the smallest and the largest mass, where 
$N$ is the number of background events determined from the sidebands.

The resulting masses\cite{conf0005} are:
\begin{eqnarray}
         m(\Bz) &=& 5279.0 \pm 0.8 \pm0.8 \mev/c^2\nonumber\\
         m(\Bpm)&=& 5278.8 \pm 0.6 \pm0.4 \mev/c^2\nonumber
\end{eqnarray}
where the first error is the quadratic sum of the statistical and uncorrelated systematic
errors and the second error is the correlated systematic error.

The mass difference between \Bz\ and \Bpm\ mesons is evaluated by fitting the \mes\ 
distributions of the three above-mentioned channels. The use of \mes\ has the advantage
that it reduces the uncertainty in the momentum scale, whilst the uncertainty due to the
beam energy cancels in the difference.  The mass difference is determined\cite{conf0005} to be:
\begin{eqnarray*}
         m(\Bz)-m(\Bpm) &=& 0.28  \pm 0.21 \pm 0.04 \mev/c^2.
\end{eqnarray*}

\section{Angular analysis of \B\to\jpsi\Kstar}

The decay of \B\to\jpsi\Kstar\ proceeds through two \CP\ even amplitudes ($A_0, A_\parallel$)
and one \CP\ odd amplitude ($A_\perp$). The contribution of $A_\perp$ must
be known before a value of \stwob\ can be determined from this decay channel.

The relative contributions of the three amplitudes are determined 
using an unbinned extended likelihood fit to the decay angles, imposing 
the constraint $|A_0|^2+|A_\parallel|^2+|A_\perp|^2=1$. The \CP\ odd fraction
is found to be $|A_\perp|^2= 0.13\pm0.06\pm0.02$ whereas
the longitudinal polarization $\Gamma_L/\Gamma$ is given by
$|A_0|^2= 0.60\pm0.06\pm0.04$.
Sources of systematic uncertainties include 
the knowledge of the background,
the acceptance corrections,
the cross-feed amongst \B\to\jpsi\Kstar\ modes
and the contribution from heavier \Kstar\ mesons.

\end{document}

%% file: pubboard/babarsym.tex
\def\lbabar{\mbox{{\large\sl B}\hspace{-0.4em} {\normalsize\sl A}\hspace{-0.03em}{\large\sl B}\hspace{-0.4em} {\normalsize\sl A\hspace{-0.02em}R}}}
\usepackage{relsize}
\def\babar{\mbox{\slshape B\kern-0.1em{\smaller A}\kern-0.1em
    B\kern-0.1em{\smaller A\kern-0.2em R}}}

\def\epem       {\ensuremath{e^+e^-}}

\def\mumu       {\ensuremath{\mu^+\mu^-}}

\def\piz   {\ensuremath{\pi^0}}

\def\pip   {\ensuremath{\pi^+}}
\def\pim   {\ensuremath{\pi^-}}
\def\pipi  {\ensuremath{\pi^+\pi^-}}

\def\Kbar  {\kern 0.2em\overline{\kern -0.2em K}{}}

\def\Kpm   {\ensuremath{K^\pm}}
\def\Kp    {\ensuremath{K^+}}

\def\Kz    {\ensuremath{K^0}}
\def\KS    {\ensuremath{K^0_{\scriptscriptstyle S}}} 
 
\def\Kstarz  {\ensuremath{K^{*0}}}

\def\Kstar   {\ensuremath{K^*}}

\def\Kstarp   {\ensuremath{K^{*+}}}

\def\Kstarpm   {\ensuremath{K^{*\pm}}}
\def\Kzb   {\ensuremath{\Kbar^0}}
\def\KzKzb {\ensuremath{K^0 \kern -0.16em \Kzb}}

\def\Dbar  {\kern 0.2em\overline{\kern -0.2em D}{}}

\def\Dzb   {\ensuremath{\Dbar^0}}
\def\DzDzb {\ensuremath{D^0 {\kern -0.16em \Dzb}}}

\def\Bz    {\ensuremath{B^0}}
\def\B     {\ensuremath{B}}

\def\Bbar  {\kern 0.18em\overline{\kern -0.18em B}{}}

\def\Bzb   {\ensuremath{\Bbar^0}}

\def\Bpm   {\ensuremath{B^\pm}}

\def\BB    {\ensuremath{B\Bbar}} 
\def\BzBzb {\ensuremath{B^0 {\kern -0.16em \Bzb}}}

\def\jpsi  {\ensuremath{{J\mskip -3mu/\mskip -2mu\psi\mskip 2mu}}} 
\def\psitwos {\ensuremath{\psi{(2S)}}}
\mathchardef\Upsilon="7107
\def\Y#1S{\ensuremath{\Upsilon{(#1S)}}}

\def\FourS {\Y4S}

\def\Xpart {\ensuremath{X}}

\mathchardef\Deltares="7101
\mathchardef\Xi="7104
\mathchardef\Lambda="7103
\mathchardef\Sigma="7106
\mathchardef\Omega="710A
\def\Deltabar   {\kern 0.25em\overline{\kern -0.25em \Deltares}{}}
\def\Lbar {\kern 0.2em\overline{\kern -0.2em\Lambda\kern 0.05em}\kern-0.05em{}}
\def\Sigbar{\kern 0.2em\overline{\kern -0.2em \Sigma}{}}
\def\Xibar{\kern 0.2em\overline{\kern -0.2em \Xi}{}}
\def\Obar{\kern 0.2em\overline{\kern -0.2em \Omega}{}}
\def\Nbar{\kern 0.2em\overline{\kern -0.2em N}{}}
\def\Xbar{\kern 0.2em\overline{\kern -0.2em X}{}}

\def\mes        {\mbox{$m_{\rm ES}$}}

\def\ev   {\ensuremath{\rm \,e\kern -0.08em V}}
\def\kev  {\ensuremath{\rm \,ke\kern -0.08em V}} 
\def\mev  {\ensuremath{\rm \,Me\kern -0.08em V}} 
\def\gev  {\ensuremath{\rm \,Ge\kern -0.08em V}} 
\def\gevc {\ensuremath{{\rm \,Ge\kern -0.08em V\!/}c}} 
\def\tev  {\ensuremath{\rm \,Te\kern -0.08em V}}
\def\mevc {\ensuremath{{\rm \,Me\kern -0.08em V\!/}c}} 
\def\gevcc{\ensuremath{{\rm \,Ge\kern -0.08em V\!/}c^2}} 
\def\mevcc{\ensuremath{{\rm \,Me\kern -0.08em V\!/}c^2}}

\def\invfb   {\ensuremath{\mbox{\,fb}^{-1}}}
\def\mus  {\ensuremath{\rm \,\mus}}

\def\mus        {\ensuremath{\,\mu{\rm s}}}    
\def\gsim{{~\raise.15em\hbox{$>$}\kern-.85em
          \lower.35em\hbox{$\sim$}~}}
\def\lsim{{~\raise.15em\hbox{$<$}\kern-.85em
          \lower.35em\hbox{$\sim$}~}}

\def\CP                 {\ensuremath{C\!P}}

\def\to                 {\ensuremath{\rightarrow}}

\def\pep2{PEP-II}

\def\chic#1{\ensuremath{\chi_{c#1}}} 

\def\stwob{\ensuremath{\sin\! 2 \beta   }}

\newcommand{\eqref}[1]{Eq.~(\ref{eq:#1})}

\newcommand{\epjc}      [1]  {{Eur.\ Phys.\ Jour.\ C~{\bf #1}}}

\newcommand{\pl}        [1]  {{Phys.\ Lett.\ {\bf #1}}}      

\def\jetset74   {\mbox{\tt Jetset \hspace{-0.5em}7.\hspace{-0.2em}4}}